\documentclass[aps,pra,reprint,superscriptaddress,noeprint]{revtex4-1}

\usepackage[hidelinks]{hyperref}
\usepackage{xcolor}
\hypersetup{
    colorlinks,
    linkcolor={black},
    citecolor={blue!80!black},
    urlcolor={blue!80!black}
}
\usepackage{graphicx}
\usepackage{amsmath,braket,mathtools}

\begin{document}


\title{Holonomic Quantum Control by Coherent Optical Excitation in Diamond}


\author{Brian B. Zhou}
\author{Paul C. Jerger}
\affiliation{Institute for Molecular Engineering, University of Chicago, Chicago, Illinois, 60637, USA}

\author{V. O. Shkolnikov}
\affiliation{Department of Physics, University of Konstanz, D-78457 Konstanz, Germany}

\author{F. Joseph Heremans}
\affiliation{Institute for Molecular Engineering, University of Chicago, Chicago, Illinois, 60637, USA}
\affiliation{Materials Science Division, Argonne National Laboratory, Argonne, Illinois 60439, USA}

\author{Guido Burkard}
\affiliation{Department of Physics, University of Konstanz, D-78457 Konstanz, Germany}

\author{David D. Awschalom}
\email{awsch@uchicago.edu}
\affiliation{Institute for Molecular Engineering, University of Chicago, Chicago, Illinois, 60637, USA}
\affiliation{Materials Science Division, Argonne National Laboratory, Argonne, Illinois 60439, USA}


\begin{abstract}
Although geometric phases in quantum evolution were historically overlooked, their active control now stimulates strategies for constructing robust quantum technologies. Here, we demonstrate arbitrary single-qubit holonomic gates from a single cycle of non-adiabatic evolution, eliminating the need to concatenate two separate cycles. Our method varies the amplitude, phase, and detuning of a two-tone optical field to control the non-Abelian geometric phase acquired by a nitrogen-vacancy center in diamond over a coherent excitation cycle. We demonstrate the enhanced robustness of detuned gates to excited-state decoherence and provide insights for optimizing fast holonomic control in dissipative quantum systems.
\end{abstract}

\pacs{}

\maketitle

Besides its central role in the understanding of contemporary physics \cite{Wilczek1989,Xiao2010}, the quantum geometric phase is gaining recognition as a powerful resource for practical applications using quantum systems \cite{Zanardi1999a,Ledbetter2012,Martin-Martinez2013}. The manipulation of nanoscale systems has progressed rapidly towards realizing quantum-enhanced information processing and sensing, but also revealed the necessity for new methods to combat noise and decoherence \cite{Lidar1998,Knill2000,Nayak2008}. Due to their intrinsic tolerance to local fluctuations \cite{Berger2013,Yale2016}, geometric phases offer an attractive route for implementing high-fidelity quantum logic. This approach, termed holonomic quantum computation (HQC) \cite{Zanardi1999a,Ekert2000,Duan2001b,Faoro2003,Zhu2002,Sjoqvist2015a}, employs the cyclic evolution of quantum states and derives its resilience from the global geometric structure of the evolution in Hilbert space. Arising both for adiabatic \cite{Berry1984} and non-adiabatic \cite{Aharonov1987} cycles, geometric phases can be either Abelian (phase shifts) or non-Abelian (matrix transformations) \cite{Wilczek1984} by acting on a single state or a subspace of states, respectively.

Recently, non-Abelian, non-adiabatic holonomic gates using three-level dynamics \cite{Sjoqvist2012b} were proposed to match the computational universality of earlier adiabatic schemes \cite{Zanardi1999a,Duan2001b,Ekert2000,Faoro2003}, but also eliminate the restriction of slow evolution. By reducing the run-time of holonomic gates, and thus their exposure to decoherence, this advance enabled experimental demonstration of HQC in superconducting qubits \cite{Abdumalikov2013}, nuclear spin ensembles in liquid \cite{Feng2013}, and nitrogen-vacancy (NV) centers in diamond \cite{Zu2014,Arroyo-Camejo2014}. However, these initial demonstrations were limited to fixed rotation angles about arbitrary axes, and thus required two non-adiabatic loops of evolution, from two iterations of experimental control, to execute an arbitrary gate \cite{Abdumalikov2013,Feng2013,Zu2014,Arroyo-Camejo2014}. Alternatively, variable angle rotations from a single non-adiabatic loop can be achieved using Abelian geometric phases \cite{Zhu2002,Wang2016a} or hyperbolic secant pulses \cite{Economou2007,Pei2010,Kodriano2012}, but these approaches are complicated by a concomitant dynamic phase. To address these shortcomings, non-Abelian, non-adiabatic single-loop schemes \cite{Xu2015,Sjoqvist2016} were designed to allow purely geometric, arbitrary angle rotations about arbitrary axes with a single experimental iteration.

In this Letter, we realize single-loop, single-qubit holonomic gates by implementing the proposal of Ref. \cite{Sjoqvist2016} in a lambda ($\Lambda$) system formed by optical transitions in the NV center. Our approach controls the common detuning and the relative amplitude and phase of a two-tone optical field that drives two non-degenerate transitions of the $\Lambda$ system. By working with detuned optical driving to an excited state rather than with resonant microwaves within the ground state \cite{Zu2014,Arroyo-Camejo2014}, we not only provide single-cycle operation and enhanced spatial resolution, but also characterize how decoherence affects gate operation. We perform quantum process tomography on an overcomplete set of resonant and off-resonant gates to demonstrate superior fidelities for off-resonant gates due to decreased excitation to the lossy intermediate state. This reveals that the detuned, single-cycle gates can offer an advantage beyond simply eliminating one pulse in the equivalent gate by two resonant pulses.

\begin{figure}[t]
 \includegraphics[scale=1]{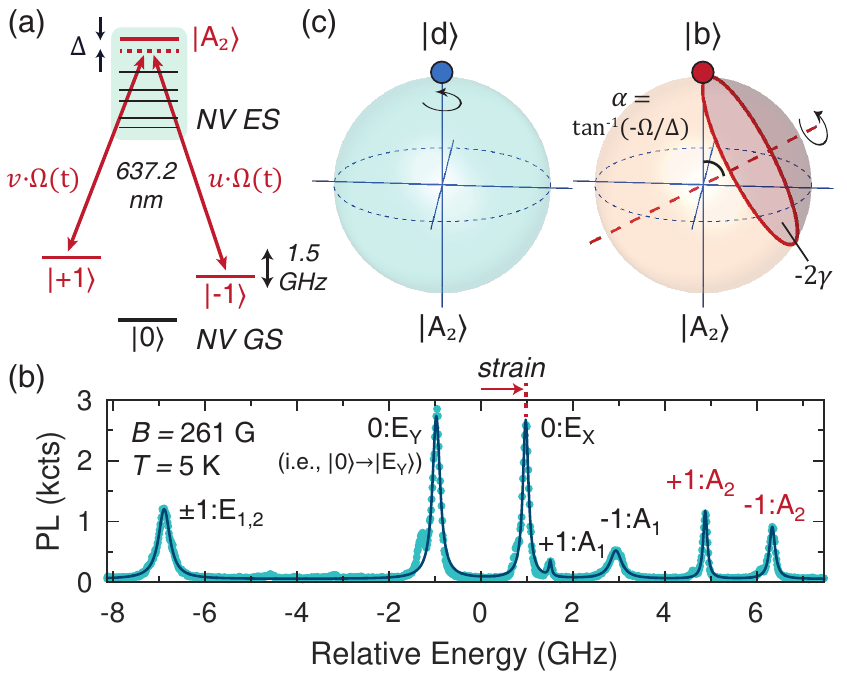}
 \caption{\label{fig:1}Experimental system and holonomic concept. (a) Optical NV $\Lambda$ system. The $\ket{\pm1}$ spin states of the NV triplet ground state (GS) are linked to $\ket{A_2}$ within the spin-orbit excited-state (ES) manifold by a two-tone optical field with one-photon detuning $\Delta$ and strength $\Omega(t)$. (b) Photoluminescence excitation spectrum taken by scanning a single laser frequency across the GS to ES transitions while two microwave tones mix the population among the three GS levels. (c) Geometric interpretation of the holonomic gates. The dark state $\ket{d}$ undergoes trivial dynamics, while the bright state $\ket{b}$ undergoes precession around a tilted axis with angle $\alpha$ on the $\ket{b}/\ket{A_2}$ Bloch sphere. After one non-adiabatic cycle, $\ket{b}$ acquires a geometric phase $\gamma$ proportional to the enclosed solid angle.}
 \end{figure}

\paragraph*{Methods.\textemdash}
We utilize a naturally-occurring, single NV center in bulk diamond cooled to 5 K. Below 20~K, optical transitions from the NV spin-triplet ground state to its spin-triplet excited state resolve narrow lines corresponding to the fine structure of the orbital-doublet excited state (Fig. \ref{fig:1}a,b) \cite{Batalov2009a}. Spin-spin and spin-orbit interactions lead to optical selection rules that enable spin-photon entanglement, as well as cycling transitions and $\Lambda$ energy structures, establishing the NV center as a leading platform for quantum optics and communication demonstrations \cite{Yale2016,Togan2010,Hensen2015,Yang2016,Zhou2016}. Here, we connect the $\ket{m_S = -1}$ and $\ket{m_S = +1}$ ground states to the highest-lying excited state $\ket{A_2}$ via a single tunable laser that is electro-optically modulated to generate frequency sidebands and nanosecond pulses \cite{Zhou2016}. We split the $\ket{\pm 1}$ states by 1.461 GHz (261 G magnetic field) and match the sideband separation to this splitting to simultaneously address both transitions, as identified in red in the photoluminescence excitation scan (Fig. \ref{fig:1}b). Due to low strain of 0.97 GHz, the three connected levels form a nearly ideal $\Lambda$ system \cite{Togan2010}.

In the rotating frame, our system is described by the Hamiltonian
\begin{multline}
	H(t) = \frac{\hbar\Omega(t)}{2} \left( u \ket{A_2}\bra{-1} + v \ket{A_2}\bra{+1} + h.c. \right) \\
		+ \Delta \ket{A_2}\bra{A_2}
\end{multline}
where $\Omega(t)$ describes the pulse envelope common to both tones and $\Delta$ is the one-photon detuning (Fig. \ref{fig:1}a). The individual transition amplitudes are scaled by the complex constants $u = \sin\left( \frac{\theta}{2}\right)$ and $v = -\cos\left( \frac{\theta}{2} \right) e^{-i\phi}$, which are controlled by tuning the relative strength and phase between the carrier and sideband frequencies. Due to the condition of two-photon resonance, this Hamiltonian admits a dark state $\ket{d} = \cos\left( \frac{\theta}{2} \right) \ket{-1} + \sin\left( \frac{\theta}{2} \right) e^{i\phi} \ket{+1}$ that is decoupled from the dynamics, and a bright state $\ket{b} = \sin\left( \frac{\theta}{2} \right) \ket{-1} - \cos\left( \frac{\theta}{2} \right) e^{i\phi} \ket{+1}$ which undergoes excitation to $\ket{A_2}$. When $\Omega(t)$ is a square pulse (i.e., $\Omega(t) = \Omega$ for $0\le t\le \tau$), $H(t)$ is time-independent during the dynamics. Thus, the expected value of the energy is conserved, remaining zero during the evolution for any initial state starting in the subspace spanned by $\ket{\pm 1}$ and ensuring the absence of dynamic phase. However, for the pulse duration $\tau = 2\pi / \sqrt{\Omega^2 + \Delta^2} \equiv T_{2\pi}$, this computational subspace undergoes cyclic, non-adiabatic evolution and transforms via the purely geometric evolution operator
$U(\theta,\phi,\Delta/\Omega) = \ket{d}\bra{d} + e^{i\gamma}\ket{b}\bra{b} =
e^{i\gamma/2}e^{-i(\gamma/2) \mathbf{n}\cdot\boldsymbol{\sigma}}$ \cite{Sjoqvist2016}, where
$\mathbf{n} = (\sin\theta\cos\phi,\sin\theta\sin\phi,\cos\theta)$,
$\boldsymbol{\sigma}$ are the Pauli matrices, and
\begin{equation}\label{eq:gamma}
\gamma = \pi \left(1 - \Delta / \sqrt{\Omega^2 + \Delta^2} \right).
\end{equation}
Up to a global phase, $U(\theta,\phi,\Delta/\Omega)$ represents a rotation by angle $\gamma$ about the axis $\mathbf{n}$ in the Bloch sphere with poles $\ket{z}=\ket{-1}$ and $\ket{-z} = \ket{+1}$, thus realizing arbitrary, non-commuting gates in a single cycle. Geometric insight is obtained by considering $\gamma$ to be the Abelian geometric phase acquired by the bright state $\ket{b}$ and equal to $-A/2$, where $A$ is the solid angle traced by $\ket{b(t)}$'s non-adiabatic precession (Fig. \ref{fig:1}c) \cite{Zu2014,Sjoqvist2016}.

Recently, Ref. \cite{Sekiguchi2017} presented an investigation of the described scheme using the $\ket{A_2} \Lambda$ system at zero field. There, a single-tone optical field addresses both degenerate transitions, while its polarization determines the bright state superposition of $\ket{\pm 1}$ that couples to $\ket{A_2}$ owing to special polarization-dependent selection rules. Rather than using this elegant but atypical correspondence between polarization and qubit states, our application instead employs a two-tone field that is generalizable to widespread non-degenerate three-level systems \cite{Kasevich1992,Hau1999,Abdumalikov2013,Pingault2014a,Rogers2014a,Safavi-Naeini2011,Dong2012}. Additionally, we perform process tomography on a universal set of single-qubit gates, including Hadamard and off-resonant gates, and identify strategies to minimize decoherence during holonomic control.

\begin{figure}[t]
 \includegraphics[scale=1]{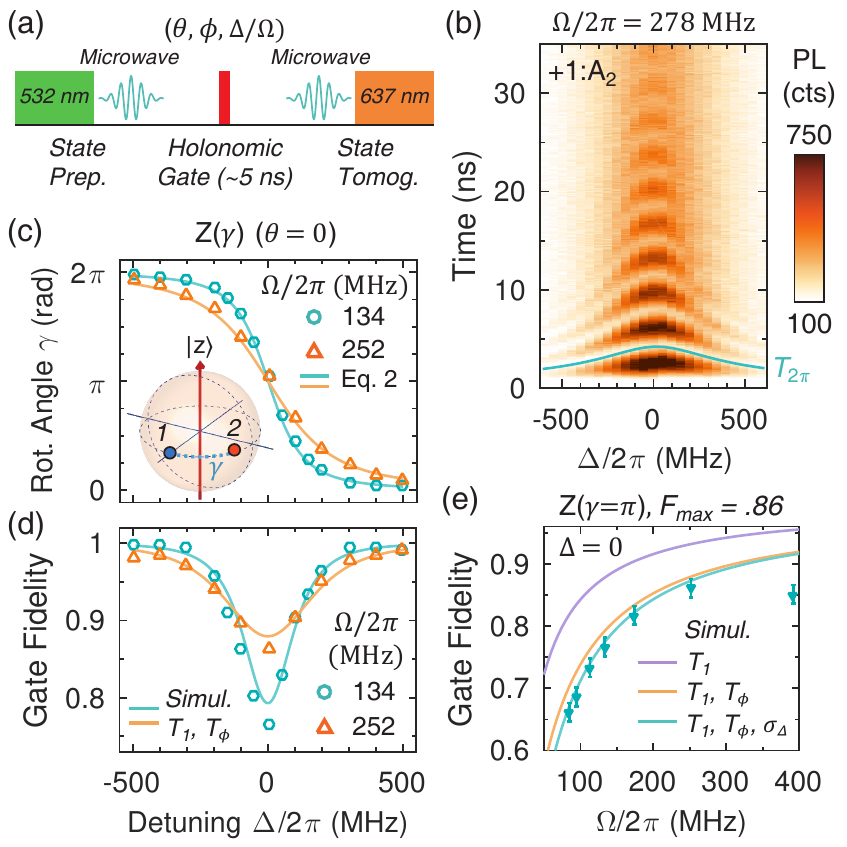}
 \caption{\label{fig:2}Phase shift gates $Z(\gamma)$ (a) Experimental sequence consisting of state preparation, optical excitation with control parameters $(\theta, \phi, \Delta/\Omega)$, and state tomography. (b) Time-resolved photoluminescence, proportional to $\ket{A_2}$ population, as a function of detuning $\Delta$, showing oscillations between $\ket{b} = \ket{+1}$ and $\ket{A_2}$ $(\theta = 0)$. (c) Measured phase shifts $\gamma$, averaged over $\ket{x}$ and $\ket{y}$ input states, as a function of $\Delta$ for $Z(\gamma)$ at two different optical powers. Solid lines delineate the prediction according to Eq. \ref{eq:gamma}. (d) Gate fidelities via process tomography of the same $Z(\gamma)$ gates. Solid lines are simulated fidelities incorporating an excited-state lifetime $T_1$ and dephasing $T_\phi$. (e) Dependence of the fidelity of the resonant gate $Z(\gamma=\pi)$ on $\Omega$. The purple, orange, and teal lines represent the simulated fidelities by sequentially adding the effects of $T_1$, $T_\phi$, and spectral hopping $\sigma_\Delta$.}
 \end{figure}

Our experiments begin by initializing the standard states $\ket{x}$, $\ket{y}$, $\ket{\pm z}$ on the $\ket{\pm 1}$ Bloch sphere (Fig. \ref{fig:2}a) \cite{Note1}. Subsequently, a short laser pulse with parameters $(\theta,\phi,\Delta/\Omega)$ implements the appropriate holonomic gate, and the resulting state is characterized by tomography along the same basis set. To determine the optical coupling $\Omega$ (in units of Rabi frequency) and the time $T_{2\pi}$ of a single excitation cycle, we measure the time-resolved photoluminescence, due to spontaneous emission from $\ket{A_2}$, during a continuous square pulse with the same $(\theta,\phi,\Delta/\Omega)$. Figure \ref{fig:2}b shows typical coherent oscillations of the population between the bright state $\ket{b}=\ket{+1}$ and $\ket{A_2}$ as a function of detuning for a drive field on the $\ket{+1}$ to $\ket{A_2}$ transition $(\theta=0)$.

\paragraph*{Results.\textemdash}
We start by characterizing arbitrary rotations about the $z$-axis via the gate set $Z(\gamma)$ with $\theta=0$. This family includes the widely-used phase shift operations Pauli-$Z$, $S$, and $T$, corresponding to $\gamma=\pi$, $\pi/2$, and $\pi/4$, respectively. In Fig. \ref{fig:2}c, we measure the phase shift $\gamma$ acquired as a function of detuning $\Delta$ for the $\ket{x}$ and $\ket{y}$ input states. We extract $\gamma$ from the difference between the phases $\varphi$ of the input and output states, where $\varphi=\tan^{-1}(Y_{p}/X_{p})$ using the Bloch vector projections $X_{p}$ ($Y_{p}$) along the $x$ ($y$)-axis. The data for two different $\Omega$ show good agreement with Eq. (\ref{eq:gamma}), delineated by the solid lines.

Using quantum process tomography, we demonstrate that the off-resonant $Z(\gamma \neq \pi)$ gates obtain higher fidelities than the resonant gate, Pauli-$Z$ (Fig. \ref{fig:2}d). Since the dominant loss mechanisms stem from $\ket{A_2}$, detuned driving reduces this detrimental exposure by decreasing both the maximal excitation and gate operation time. The solid lines in Fig. \ref{fig:2}d represent the expected fidelities from a master equation simulation \footnote{For additional details, see Supplemental Material, which includes Refs. \cite{Altepeter2004} and \cite{Howard2006}} that uses an excited-state lifetime ($T_1=11.1$ ns) and dephasing ($T_\phi=18$ ns) as previously measured for this NV center \cite{Zhou2016}. Furthermore, we demonstrate that the fidelity of the resonant Pauli-$Z$ gate increases as the run-time $T_{2\pi} \propto \Omega^{-1}$ decreases (Fig. \ref{fig:2}e) \cite{Johansson2012}. We separate the theoretical contributions to decoherence due to the lifetime $T_1$ and dephasing $T_\phi$ by incorporating their effects sequentially. The remaining discrepancy with the data, particularly at low optical powers, is largely reconciled by introducing spectral hopping of the excited state's energy, modeled by detuning errors with a Gaussian standard deviation $\sigma_\Delta =2\pi\cdot15$ MHz. The experimental fidelities saturate at $F = .86(2)$ for $\Omega/2\pi > 252$ MHz due to crosstalk with nearby levels (Fig. \ref{fig:1}b) and laser leakage before and after the pulse caused by a finite extinction ratio \cite{Note1}.

\begin{figure*}[t]
 \includegraphics[scale=1]{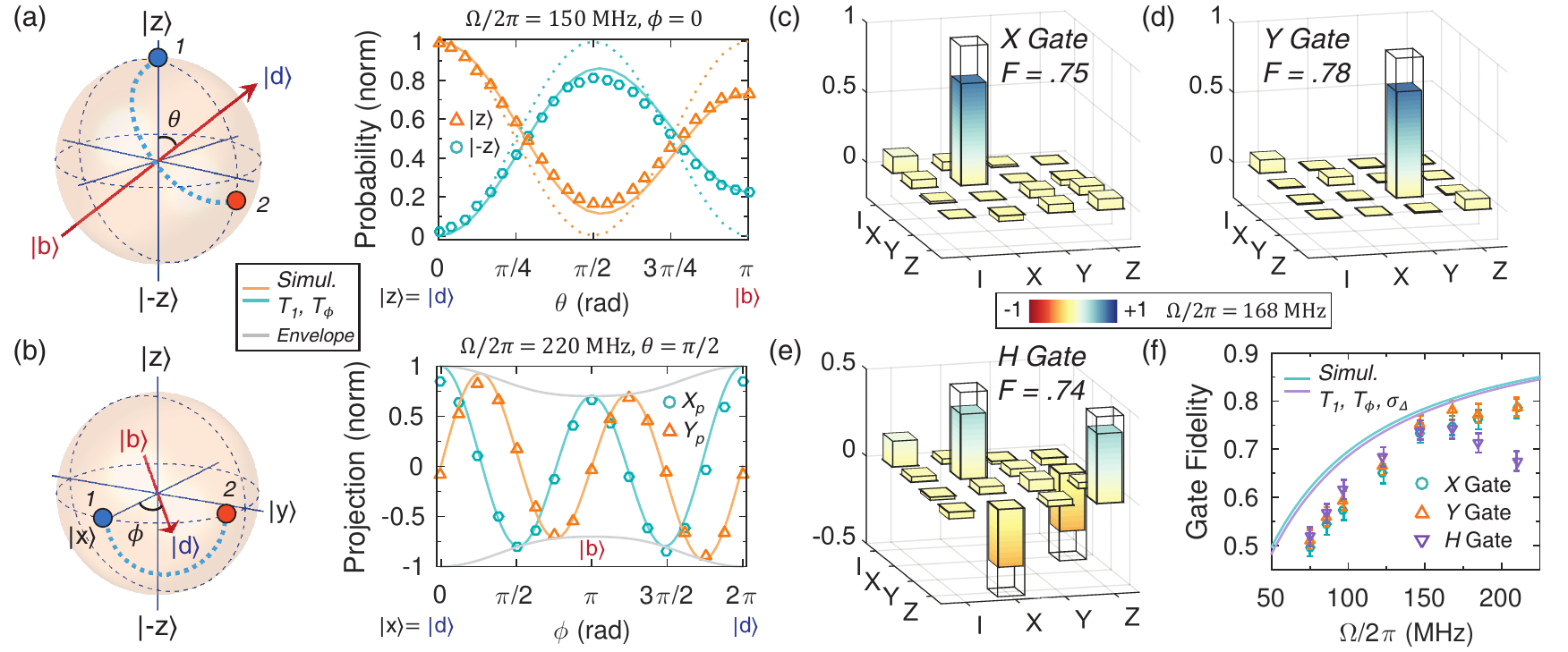}
 \caption{\label{fig:3}Resonant holonomic gates.  (a) Gates with variable $\theta$ after initializing $\ket{z}$. The probabilities of the final states to be measured in $\ket{z}$ and $\ket{-z}$ are plotted. (b) Gates with variable $\phi$ after initializing $\ket{x}$. The Bloch vector projections of the final states along the $x$- and $y$-axes are plotted. For both (a) and (b), solid lines indicate simulated behaviors with $T_1$ and $T_\phi$ decoherence; dashed lines in (a) indicate decoherence-free behavior. Decoherence is minimized when the initialized state aligns with the gate's dark state $\ket{d}$. (c,d,e) Process matrices for $X$, $Y$, and $H$ at $\Omega/2\pi = 168$ MHz. (f) Dependence of $X$, $Y$, $H$ process fidelities on $\Omega$. The solid lines are simulated fidelities incorporating all excited-state decoherence effects. $X$ and $Y$ are identical in the simulation.}
\end{figure*}

We proceed to examine resonant holonomic $\pi$-rotations about arbitrary axes in Fig. \ref{fig:3}. Before focusing on process tomography of the Pauli-$X$, Pauli-$Y$, and Hadamard gates (denoted $X$, $Y$, $H$, respectively \cite{Note1}), we verify full control over the $(\theta,\phi)$ degrees of freedom and illuminate how the relationship between the gate's input state and dark/bright state axis affects decoherence. In Fig. \ref{fig:3}a, we initialize $\ket{z}$ and apply gates with variable $\theta$, holding $\phi=0$. As schematically illustrated, this holonomic transformation induces effective Rabi oscillations between $\ket{z}$ and $\ket{-z}$ as $\theta$ is increased. Notably, the discrepancy between the realized population transfer (data points) and the decoherence-free transfer (dashed lines) varies as a function of $\theta$, being smallest (largest) at $\theta=0$ $(\theta=\pi)$ when the initialized state $\ket{z}$ is the dark (bright) state of the gate (Fig. \ref{fig:3}a). Correspondingly, we initialize $\ket{x}$ and apply gates with variable $\phi$, holding $\theta=\pi/2$ (Fig. \ref{fig:3}b). In this case, we display $X_{p}$ and $Y_{p}$ of the final state, showing it to rotate around the equator twice as $\phi$ rotates once, as expected. The visibility of the final state (gray envelope) is maximal (minimal) when $\ket{x}$ is aligned with the gate's dark (bright) state. Since the on-resonant gates ($\pi$-rotations) are invariant under interchange of $\ket{d}$ and $\ket{b}$, there exist two equivalent choices $\{\theta, \theta+\pi\}$ for each gate in the absence of dissipation. However, for implementations where the intermediate state dominates loss, our observations demonstrate that when the input state $\ket{\psi_i}$ is known (e.g., in state preparation), the more effective holonomic gate minimizes $\left| \braket{\psi_i|b} \right|^2$.

We achieve process fidelities for $X$, $Y$, and $H$ of .75(2), .78(2), and .74(2), respectively, at $\Omega/2\pi=168$ MHz (Fig. \ref{fig:3}c-e). The fidelities of these gates display a similar scaling versus $\Omega$ as the $Z$ gate. However, when comparing with the simulated fidelities (solid lines) using the same excited-state $T_1$, $T_\phi$, and $\sigma_\Delta$, the data for $X$, $Y$, and $H$ realize lower fidelities than expected by our model (Fig. \ref{fig:3}f). This is explained by the presence of additional experimental non-idealities from turning on two drive frequencies, such as enhanced crosstalk, relative phase errors, and extraneous higher-harmonic sidebands. We improve the fidelity of $X$ and $Y$ to .79(2) at $\Omega/2\pi=210$ MHz, but $H$ decreases in fidelity. This effect may stem from the greater susceptibility of $H$ to systematic errors.

\begin{figure}[t]
 \includegraphics[scale=1]{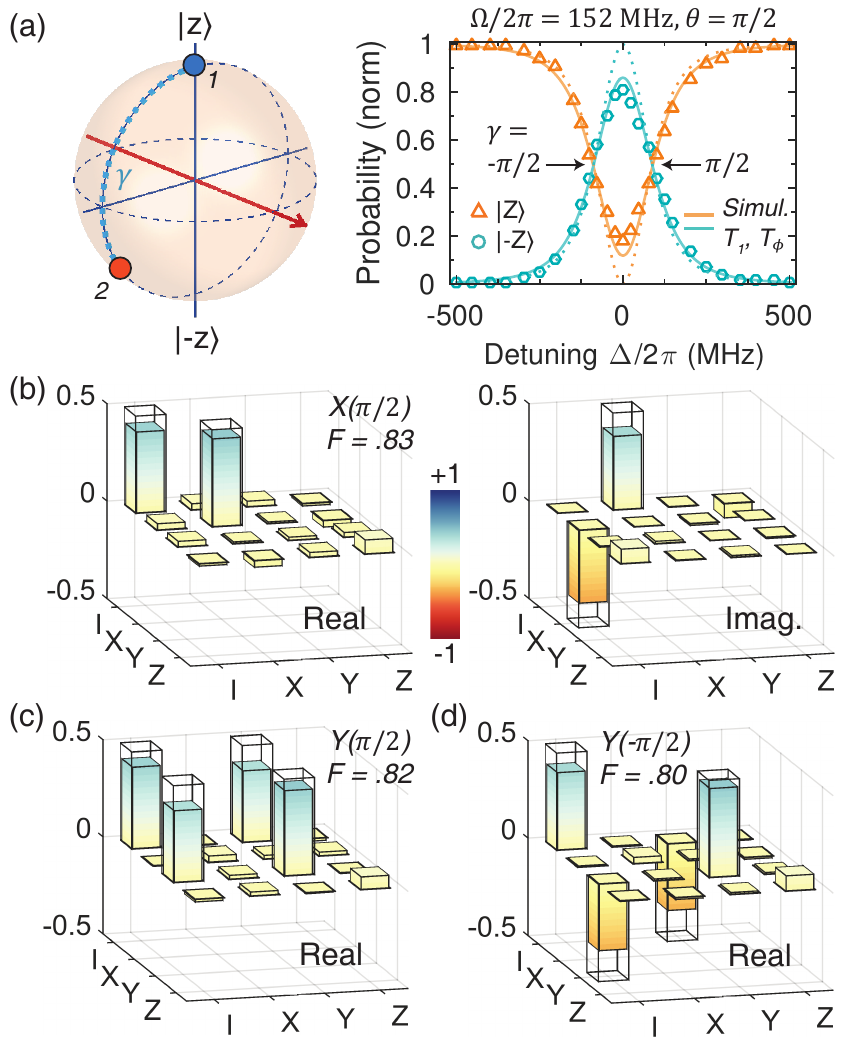}
 \caption{\label{fig:4}Variable rotations around $x$- and $y$-axes. (a) After initializing $\ket{z}$, a holonomic gate with variable detuning $\Delta$ and fixed rotation axis along $\ket{x}$ is applied. The data show Rabi oscillations between $\ket{z}$ and $\ket{-z}$ as $\Delta$ is swept across resonance, tuning the rotation angle $\gamma$. Solid lines are simulations with $T_1$ and $T_\phi$ decoherence; dashed lines are decoherence-free behaviors. (b,c,d) Process matrices for the $X(\pi/2)$ and $Y(\pm\pi/2)$ gates at $\Omega/2\pi = 152$ MHz and $\Delta/2\pi = \pm 88$ MHz.}
 \end{figure}

Finally, we demonstrate tunable rotations around the $x$- and $y$-axes by varying the detuning of the optical pulse. Initializing $\ket{z}$, we verify that these gates rotate the population from $\ket{z}$ to $\ket{-z}$ and back as the laser frequency is varied across one-photon resonance (Fig. \ref{fig:4}a, $\theta = \pi/2$). Focusing on $\Delta/\Omega=\pm1/\sqrt{3}$ for $\Omega/2\pi = 152$ MHz, we realize process fidelities for the $X(\pi/2)$ (i.e., $\sqrt{NOT}$ gate), $X(-\pi/2)$, $Y(\pi/2)$, and $Y(-\pi/2)$ gates of .83(2), .80(2), .82(2), and .80(2), respectively, where the sign of the rotation is controlled by the sign of $\Delta$ (Fig. \ref{fig:4}b-d) \cite{Note1}. Due to decreased excitation, these gate fidelities exceed those of the resonant gates at the same optical power ($\sim$.74 at $\Omega/2\pi =150$ MHz). In the resonant scheme for non-adiabatic HQC \cite{Sjoqvist2012b}, arbitrary angle rotations required two $\pi$-rotations around different axes: for example, $Y(\pi/2)=X\cdot H$. Our data demonstrate that in applications involving dissipative intermediate states, the single-loop scheme significantly outperforms the equivalent composite gate, which would achieve here an estimated fidelity of .55 ($\approx .74^2$), and moreover exceeds the fidelity of a \emph{single} resonant gate. In comparison, if the single-loop gates are implemented by microwave driving within the ground state \cite{Zu2014,Arroyo-Camejo2014}, the intermediate level $\ket{m_S=0}$ would decohere at a rate (limited by ground-state dephasing $T_2^* \sim$ 10 $\mu s$) comparable to the computational states $\ket{\pm1}$. This extended coherence enables higher fidelities, but decreased occupation of $\ket{0}$ by off-resonant driving would not be an advantage beyond its reduction in the number of gates applied.

\paragraph*{Discussion.\textemdash}
The optical gate fidelities are limited by excited-state occupation and crosstalk between the driving fields. Improving the rise time and extinction of our optical pulses would enable faster transit through the excited state without side effects and also reduce errors due to dynamic phase for the detuned gates, which are fully geometric only for perfect square pulses \cite{Sjoqvist2016}. Although for $\Omega/2\pi = 152$ MHz, the current rise/fall times ($\sim$1.2 ns) contribute negligible errors to $X(\pi/2)$, we estimate that dynamic errors would begin to limit the fidelity for $\Omega/2\pi > 600$ MHz \cite{Note1}. Crosstalk between the $\Lambda$ transitions can be improved by using orthogonal polarizations for the two driving frequencies, rather than the same linear polarization, to exploit the polarization-selectivity of the transitions. In Ref. \cite{Sekiguchi2017}, faster gate speeds and reduced crosstalk in the single-tone implementation enabled fidelities of $\sim$.90 for the resonant Pauli gates, while calculations are required to compensate the driving polarization for each specific NV strain and orientation.

Our experiments establish universal single-qubit holonomic control of solid-state spins with optical spatial resolution and single-cycle operation. A path toward two-qubit gates and universal computation is envisioned by leveraging strong-coupling to nearby nuclear spins \cite{Jelezko2004} or cavity-mediated interactions to other NVs \cite{Burkard2014,Zhou2017a}. Fundamentally, our holonomic operations are efficient in both time and number of control parameters for the arbitrary manipulation of qubit levels that are not directly coupled. Thus, they are relevant to hybrid systems where disparate quantum systems are indirectly linked via an intermediary. Holonomic state transfer may be more efficient than sequential, multi-pulse operations through the dissipative intermediate system \cite{Wang2012a}. Additionally, our methods offer an alternative when far-detuned stimulated Raman transitions $(\Delta\gg\Omega)$, which effect much slower rotations, are impractical due to the interplay between decoherence rates and level separations \cite{Yale2013a,Golter2014a,Childress2014}. The strategies demonstrated here for optimizing fast holonomic control enrich the quantum control toolbox to adapt to a growing diversity of useful quantum systems.\\

\begin{acknowledgments}
We thank C. G. Yale and A. Baksic for valuable discussions. This work was supported by the Air Force Office of Scientific Research MURI FA9550-15-1-0029 and FA9550-14-1-0231, and the National Science Foundation DMR-1306300. FJH and DDA contributed to the experimental design, analysis of data, and preparation of the manuscript and were supported by the U.S. Department of Energy, Office of Science, Office of Basic Energy Sciences, Materials Sciences and Engineering Division. Work at the University of Konstanz was supported by the German Research Foundation (SFB 767).

\paragraph*{Note added.\textemdash}
During preparation of our manuscript, we became aware of two complementary works \cite{Sekiguchi2017,Li2017}.
\end{acknowledgments}

\clearpage
\setcounter{equation}{0}
\setcounter{figure}{0}
\onecolumngrid

\renewcommand{\thesection}{S\arabic{section}}
\renewcommand{\thefigure}{S\arabic{figure}}
\renewcommand{\theequation}{S\arabic{equation}}
\renewcommand*{\citenumfont}[1]{S#1}
\renewcommand*{\bibnumfmt}[1]{[S#1]}


\begin{center}
{\fontsize{12}{12}\selectfont
\textbf{Supplemental Material for
	``Holonomic Quantum Control by Coherent Optical Excitation in Diamond''\\ [5mm]}}
{\normalsize Brian B. Zhou,\textsuperscript{1} Paul C. Jerger,\textsuperscript{1} V. O. Shkolnikov,\textsuperscript{2} F. Joseph Heremans,\textsuperscript{1,3}\\ Guido Burkard,\textsuperscript{2} and David D. Awschalom\textsuperscript{1,3}\\[2mm]}
{\fontsize{9}{9}\selectfont
\textsuperscript{1}\textit{Institute for Molecular Engineering, University of Chicago, Chicago, Illinois 60637, USA}\\[1mm]
\textsuperscript{2}\textit{Department of Physics, University of Konstanz, D-78457 Konstanz, Germany}\\[1mm]
\textsuperscript{3}\textit{Materials Science Division, Argonne National Laboratory, Argonne, Illinois 60439, USA}}
\end{center}   
\normalsize

\section{Experimental Setup\label{setup}}
The sample used in this experiment is (100)-oriented electronic-grade type IIa diamond from Element Six. A single NV center was identified 2 $\mu$m below the surface, and imaged using a custom-built confocal microscopy setup containing an objective with NA = 0.9. A static magnetic field of 261 G aligned to the NV axis splits the $\ket{m_S = \pm 1}$ ground state sublevels by 1.461 GHz. A closed-cycle cryostat (Montana Instruments) cooled the sample to 5 K, allowing observation of the 1.94 GHz ($= 2\delta_\perp$) strain splitting of the $\ket{E_X}$ and $\ket{E_Y}$ excited state levels. After a 532-nm diode laser initialized the spin state into $\ket{m_S = 0}$, resonant microwave pulses (at 2.147 and 3.608 GHz) delivered from a coplanar waveguide prepared the NV into the desired superposition state of $\ket{m_S = \pm 1}$.

Two tunable 637-nm diode lasers were incorporated with a dichroic mirror, one to resonantly read out the spin state, and the other to perform the holonomic gates. This interaction laser was passed through fiber-coupled phase (PEOM; Jenoptik PM 635) and amplitude (AEOM; Jenoptik AM 635) electro-optic modulators. The PEOM creates frequency harmonics at the driving frequency (1.461 GHz) and determines their relative amplitudes and phases, allowing the $\ket{-1}\rightarrow\ket{A_2}$ transition to be driven by the carrier frequency and the $\ket{+1}\rightarrow\ket{A_2}$ transition by the red-shifted first sideband. When the single $\ket{+1}\rightarrow\ket{A_2}$ transition was driven (Fig. 2 of main text), then the frequency of the carrier was shifted, without requiring sidebands to be generated by the PEOM. Furthermore, in this case, the polarization of the interaction laser was set to minimize coupling to the other transition. The AEOM shapes nanosecond-scale laser pulses (see Section \ref{shape}) and compensates for the PEOM modulation to maintain constant combined intensity in the two utilized harmonics as their ratio is changed. The detuning of the interaction laser was stabilized by a wavemeter (High Finesse WS6-200).

The microwaves, modulators, and lasers were controlled and coordinated by two synchronized arbitrary waveform generators (AWG; Tektronix 5014C, 1 GSample/s; and Tektronix 7102, 10 GSample/s). To read out the state of the NV center on the $\ket{m_S = \pm 1}$ Bloch sphere, appropriate microwave pulses transferred the state to $\ket{0}$ at a rephasal time of the $^{14}\rm{N}$ nuclear spin. The phonon sideband emission from the $\ket{0}\rightarrow\ket{E_Y}$ transition was then separated from the zero-phonon-line emission by filtering between 650 nm to 800 nm and detected with an avalanche photodiode (APD; PerkinElmer SPCM-AQR-16-FC). Additional details for the calibration of the AEOM and PEOM by optical Rabi oscillations, as well as measurement of excited state decay and dephasing rates for the NV used here, are supplied in the Supplementary Information of Ref. \cite{Zhou2016Supp}.

\section{Pulse Shape Analysis\label{shape}}
\begin{figure}[h]
	\centering
	\includegraphics[width=0.9\linewidth]{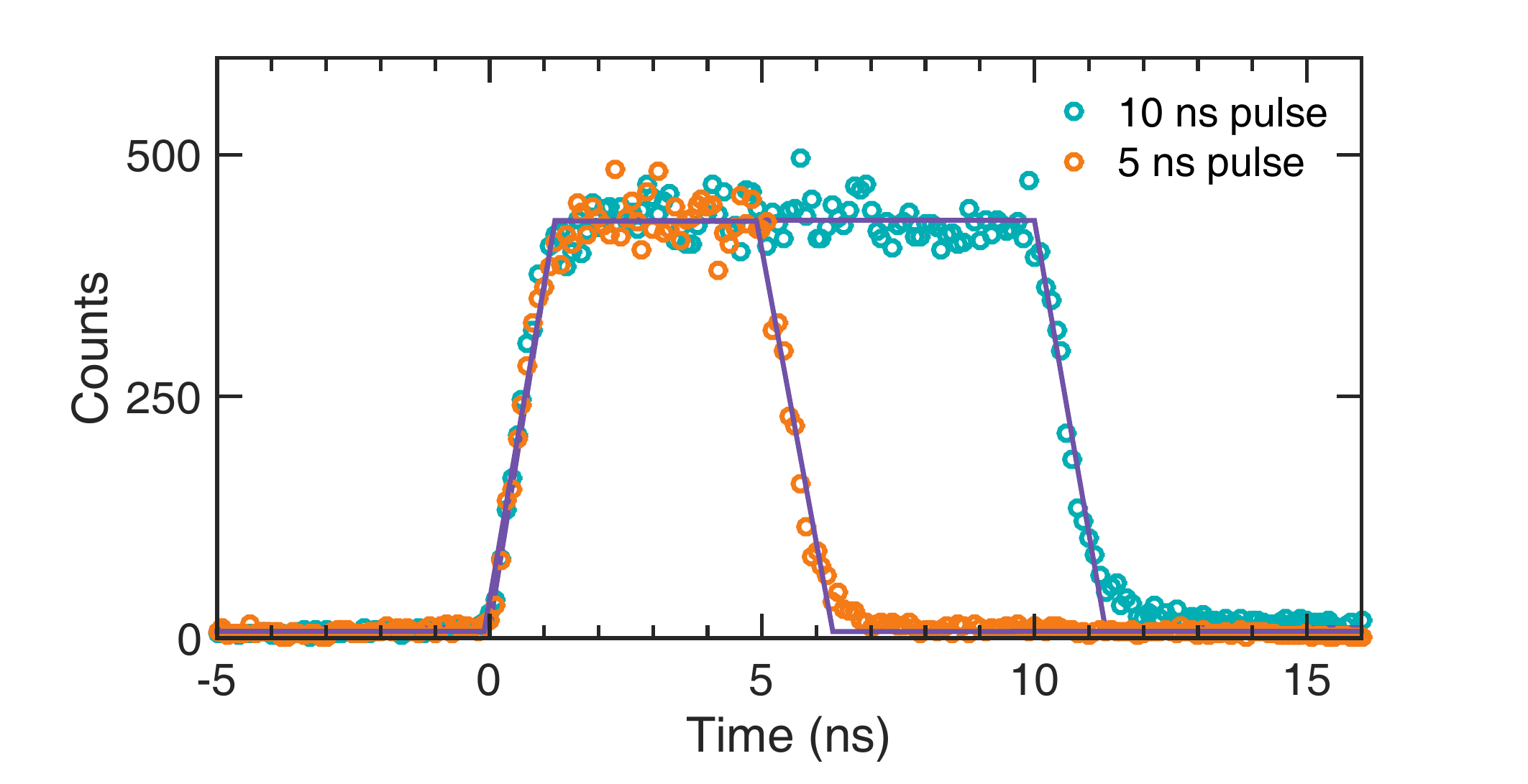}
	\caption{Pulse data for square pulses shaped with AEOM. Counts are read from pulse reflections off the diamond surface, away from NV centers. Rise/fall times are approximately 1.2 ns.}
	\label{fig:hqcsipulseshape}
\end{figure}

Since off-resonant holonomic gates in a lambda system are fully geometric only for square pulses \cite{Sjoqvist2016Supp}, we characterized our applied pulses by measuring their reflection off the sample with the APD. Fig. \ref{fig:hqcsipulseshape} shows the results of applying 5- and 10-ns square wavefroms to the AEOM. The resulting optical pulse shapes are well-approximated by a trapezoidal shape, with a best-fit 1.2$\pm$0.1 ns turn-on ramp and 1.3$\pm$0.1 ns turn-off ramp. Accordingly, a pulse fall time correction is applied when determining the cycle time $T_{2\pi}$ from optical Rabi oscillations using a continuous square pulse (e.g. Fig. 2b of the main text). All simulations use a trapezoidal pulse shape with 1.2 ns rise and fall times. Additionally, we note from the data that the extinction ratio (on/off) of our pulse is $\sim$50:1. However, a separate acousto-optic modulator, which has much higher extinction ($>$45 dB), brackets the AEOM, limiting this leakage to $\sim$10 ns before and after the pulse. See Section \ref{theory} and Fig.  \ref{Pulse_form} for discussion of pulse shape effects on the fidelity of the holonomic gates.

\newpage
\section{Parameters for Holonomic Gates\label{params}}
\setlength{\tabcolsep}{12pt}
\begin{table}[h]
	\centering
	\begin{tabular}{ccccr}
		\hline
		Gate & Loop Parameter $\theta$ & Loop Parameter $\phi$ & Detuning ($\Delta/\Omega$) & $ U(\theta,\phi,\Delta/\Omega)$ \\
		\hline
		X & $\pi/2$ & 0 & 0 & $\begin{pmatrix} \mathmakebox[\widthof{$-1$}]{0} & \mathmakebox[\widthof{$-1$}]{1} \\ 1 & 0 \end{pmatrix}$ \\
		Y & $\pi/2$ & $\pi/2$ & 0 & $\begin{pmatrix} \mathmakebox[\widthof{$-1$}]{0} & \mathmakebox[\widthof{$-1$}]{-i} \\ i & 0 \end{pmatrix}$ \\
		Z($\gamma$) & 0 & N/a & $\delta$ & $\begin{pmatrix} \mathmakebox[\widthof{$-1$}]{1} & \mathmakebox[\widthof{$-1$}]{0} \\ 0 & e^{i\gamma} \end{pmatrix}$ \\
		H & $3\pi/4$ & 0 & 0 & $\frac{1}{\sqrt{2}} \begin{pmatrix} \mathmakebox[\widthof{$-1$}]{-1} & \mathmakebox[\widthof{$-1$}]{1} \\ 1 & 1 \end{pmatrix}$ \\
		X($\pi/2$) & $\pi/2$ & 0 & $\frac{1}{\sqrt{3}}$ & $\frac{1+i}{2} \begin{pmatrix} \mathmakebox[\widthof{$-1$}]{1} & \mathmakebox[\widthof{$-1$}]{-i} \\ -i & 1 \end{pmatrix}$ \\
		X(-$\pi/2$) & $\pi/2$ & 0 & $\frac{-1}{\sqrt{3}}$ & $\frac{1-i}{2}\begin{pmatrix} \mathmakebox[\widthof{$-1$}]{1} & \mathmakebox[\widthof{$-1$}]{i} \\ i & 1 \end{pmatrix}$ \\
		Y($\pi/2$) & $\pi/2$ & $\pi/2$ & $\frac{1}{\sqrt{3}}$ & $\frac{1+i}{2} \begin{pmatrix} \mathmakebox[\widthof{$-1$}]{1} & -1 \\ 1 & 1 \end{pmatrix}$ \\
		Y(-$\pi/2$) & $\pi/2$ & $\pi/2$ & $\frac{-1}{\sqrt{3}}$ & $\frac{1-i}{2} \begin{pmatrix} 1 & \mathmakebox[\widthof{$-1$}]{1} \\ -1 & 1 \end{pmatrix}$ \\
		\hline
	\end{tabular}
	\caption{Experimental parameters for implementing various holonomic
		gates. Unless specified in parentheses, each gate performs a $\pi$ rotation about the axis $\mathbf{n} = (\sin\theta\cos\phi,\sin\theta\sin\phi,\cos\theta)$, corresponding to the holonomic transformation $U(\theta,\phi,\Delta/\Omega)$. For the Z gate, $\gamma = \pi\left(1 - \delta/\sqrt{1 + \delta^2} \right)$, where $\delta = \Delta/\Omega$. Note the phase definition on the Hadamard H gate.}
\end{table}

\section{Quantum Process Tomography\label{defs}}
As set by the alignment of the NV axis and the static magnetic field, the
$\ket{z} = \ket{m_s = -1}$ and $\ket{-z} = \ket{m_s = +1}$ spin sublevels are
the poles of the qubit Bloch sphere. We use standard definitions for the
orthogonal states:
\begin{align*}
	\ket{\pm x} &= \frac{1}{\sqrt{2}} \left( \ket{z} \pm \ket{-z} \right) \\
	\ket{\pm y} &= \frac{1}{\sqrt{2}} \left( \ket{z} \pm i\ket{-z} \right)
\end{align*}
For quantum process tomography, we prepare the input states $\ket{z}$, $\ket{-z}$, $\ket{x}$, and $\ket{y}$, apply the holonomic gate, and then perform standard quantum state tomography on the output state \cite{Altepeter2004Supp,Howard2006Supp}. This entails measuring the projection (probability) of the final state along the basis set $\ket{\pm z}$, $\ket{\pm x}$, and $\ket{\pm y}$ in order to determine the final density matrix $\rho^{final}_{exp}$. In general, the density matrix resulting from a quantum operation is characterized by a process matrix $\chi$, such that $\rho_{final}=\sum_{i,j=1}^{4}\chi_{ij} E_i\rho_{initial} E_j$. We choose the fixed set of basis operators to be $E_1 = I$, $E_2 = X = \sigma_x$, $E_3 = Y = -i \sigma_y$, and $E_4 = Z = \sigma_z$. Inverting the relation for $\rho^{final}_{exp}$, we determine the process matrix $\chi_{exp}$, which is then re-estimated by a positive semidefinite matrix parametrization to ensure physicality. However, we do not enforce $Tr(\chi) = 1$ due to the possibility of leakage out of the computational subspace into $\ket{m_s = 0}$ via weak pathways. Hence, the realized $\chi_{exp}$ have traces slightly less than one. To control for measurement errors unrelated to the holonomic gate, we perform process tomography on the identity element (i.e., same experimental sequence but without an interaction laser pulse). We define the net fidelity of the experiment, including measurement errors, to be $\mathcal{F} =  \mathrm{Tr}(\chi_{exp}\chi_{ideal})$, where $\chi_{ideal}$ is the process matrix for the ideal operation. The reported gate fidelities $F$ in the main text correspond to $F = \mathcal{F}_{gate}/ \mathcal{F}_{Identity}$. This normalizes for measurement errors, and generally $\mathcal{F}_{Identity} = .97(1)$. Errors are reported at 95\% confidence and are estimated from the statistical errors on the elements of $\rho^{final}_{exp}$.

\section{Additional Data}
\subsection{Phase Shift Gates $Z(\gamma)$}
\begin{figure}[h]
	\centering
	\includegraphics[width=1.0\textwidth]{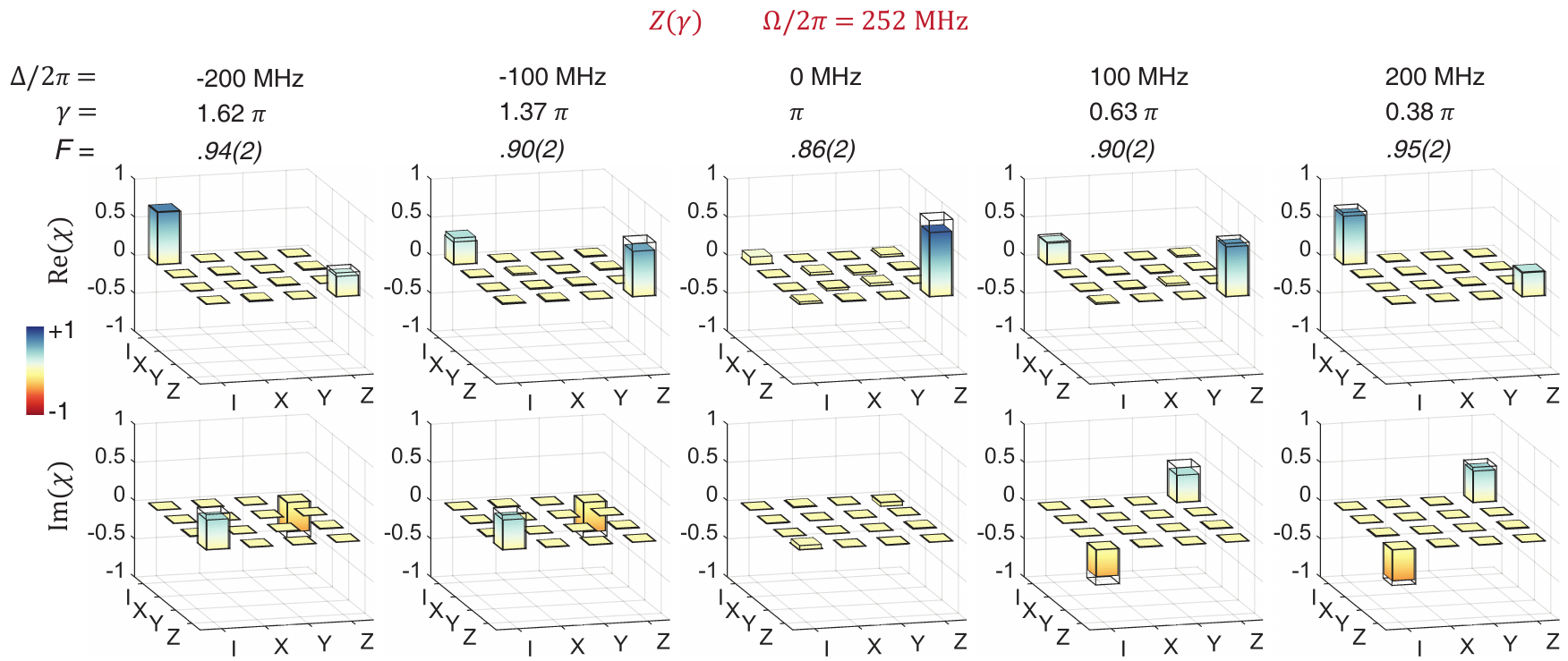}
	\caption{Example process matrices for $Z(\gamma)$, corresponding to the phase shifts and fidelities reported in Fig. 2 of the main text. The top (bottom) panel denotes the real (imaginary) part of the process matrix $\chi$, and the transparent bars denote the entries in the ideal process matrix. At $\Omega/2\pi$ = 252 MHz, the fidelity $F$ of $Z(\gamma)$ is .86(2) for a resonant $\pi$-rotation and increases for non-zero detuning $\Delta$.}
	\label{fig:z_gamma}
\end{figure}

\newpage
\subsection{Holonomic Gates with Variable $\theta$}
\begin{figure}[h]
	\centering
	\includegraphics[width=1.0\textwidth]{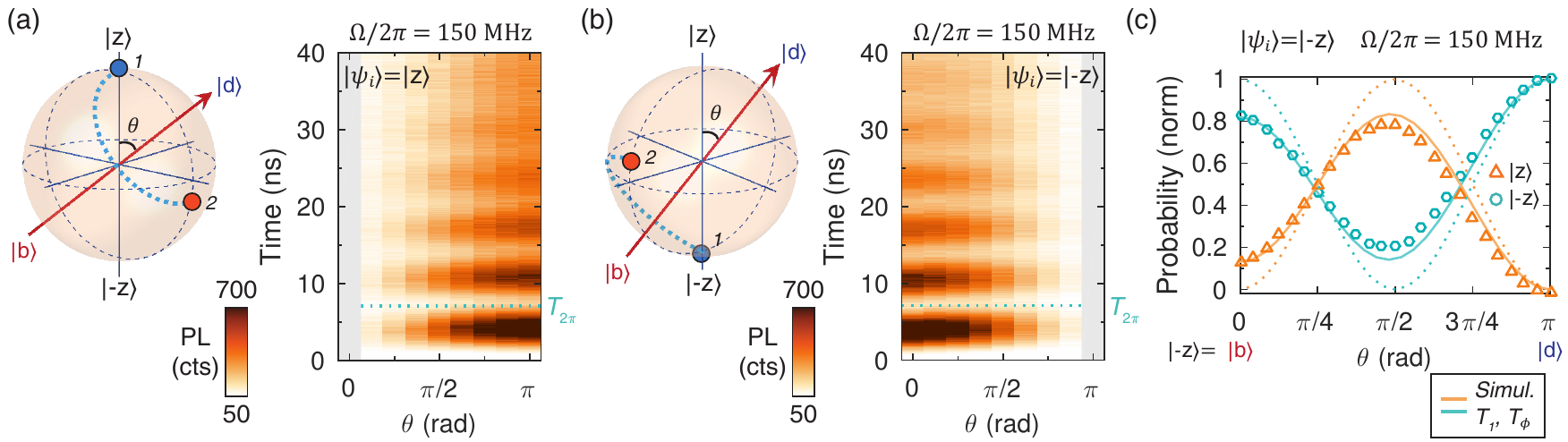}
	\caption{(a) Time-resolved photoluminescence, proportional to the occupation of $\ket{A_2}$, for gates with variable $\theta$ and $\phi = 0$ when the initialized state is $\ket{z}$. Excitation to $\ket{A_2}$ is minimized for $\theta = 0$ when $\ket{z}$ is the dark state $\ket{d}$ of the gate. (b) Same measurement after initializing the state $\ket{-z}.$ Now excitation is minimized for $\theta = \pi$, as expected. The constant cycle time $T_{2\pi}$ as a function of $\theta$ for both initial states indicates the correct calibration of our optical modulators (PEOM and AEOM). (c) Complementary data set to Fig. 3a of the main text. The probabilities of the final states to be measured in $\ket{z}$ and $\ket{-z}$ are plotted. Here, we initialize $\ket{-z}$ and show it to be transferred to $\ket{z}$ as $\theta$ increases to $\pi/2$ ($X$ gate). Solid lines are simulations with $T_1$ (excited-state lifetime) and $T_\phi$ (dephasing time) decoherence; dashed lines are decoherence-free behaviors. As in Fig. 3a of the main text, contrast in fidelity is lowest when the initialized state is the bright state, due to increased excitation to $\ket{A_2}$.}
	\label{fig:theta_rabi}
\end{figure}

\subsection{Detuned Rotations About X-Axis}
\begin{figure}[h]
	\centering
	\includegraphics[width=1.0\textwidth]{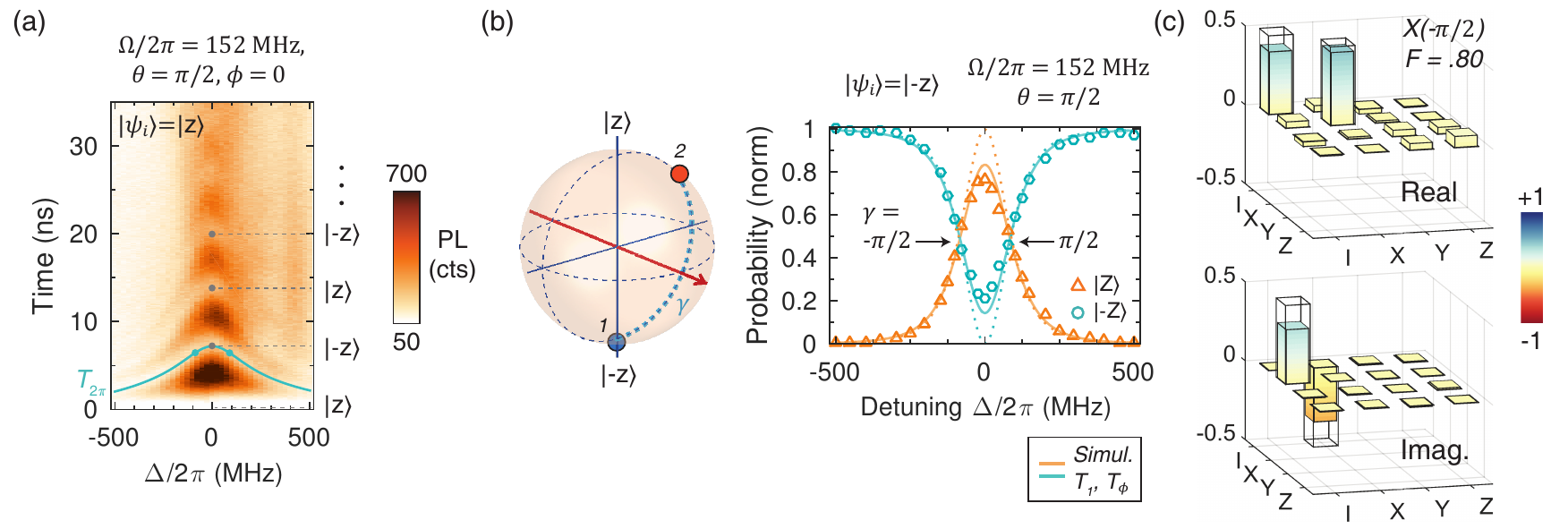}
	\caption{(a) Time-resolved photoluminescence, proportional to $\ket{A_2}$ population, as a function of detuning $\Delta$ for $\theta = \pi/2$, $\phi = 0$. The initialized state is $\ket{z}$. On resonance ($\Delta = 0$), the holonomic gate $X$ swaps $\ket{z}$ to $\ket{-z}$ and vice versa for each complete excitation cycle, at times marked by the gray points. The teal points indicate the detuning and pulse duration for executing a $\pm\pi/2$ rotation around the $x$-axis. (b) Complementary data set to Fig. 4a in the main text. Here the initialized state is $\ket{-z}$. A holonomic gate with variable detuning $\Delta$ and fixed rotation axis along $\ket{x}$ is applied. The data show Rabi oscillations between $\ket{z}$ and $\ket{-z}$ as $\Delta$ is swept across resonance, tuning the rotation angle $\gamma$. Solid lines are simulations with $T_1$ (excited-state lifetime) and $T_\phi$ (dephasing time) decoherence; dashed lines are decoherence-free behaviors. (c) Real and imaginary parts of the experimental quantum process tomography matrix for the $X(-\pi/2)$ gate with fidelity $F=0.80(2)$.}
	\label{fig:detune_data}
\end{figure}

\section{Master Equation Simulation of Holonomic Quantum gates\label{theory}}
In this section, we describe the model used to simulate the behavior of the holonomic quantum gates. We will describe the Hamiltonian and the main sources of decoherence, as well as how we calculated the theoretical fidelity of the gates and output states.
\subsection{Hamiltonian and environmental effects}
The fine structure of the NV center consists of a spin triplet in the ground state and a spin triplet, orbital doublet in the excited state. Since the ground states $ \ket{-1}$, $\ket{+1}$ are experimentally coupled to the $\ket{A_2}$ excited state by laser excitations that are close to resonance, we neglect the effect of the other five excited states. We consider the NV center as a 4-level system by including the ground state $\ket{0}$, which can be populated by weak decay paths from $\ket{A_2}$ due to the non-zero strain and intersystem crossing. We go to a rotating frame in which all three levels of the ground state have the same energy, while the excited state $\ket{A_2}$ is higher in energy by the one-photon detuning $\Delta$. The coherent part of the system dynamics is described with the following Hamiltonian
\begin{equation}
	H=\frac{\Omega(t)}{2}(\sin(\theta/2)\ket{-1}\bra{A_2}-\cos(\theta/2)e^{i\phi}\ket{+1}\bra{A_2}+h.c.)+\Delta \ket{A_2}\bra{A_2}.
\end{equation}
Here, $\theta$ and $\phi$ define the relative amplitude and phase of the two driving tones used to perform the gate, and $\Omega(t)$ is a pulse envelope common for both tones. The optical fields contain no oscillations as we work in the rotating frame. We model the laser pulse as a trapezoid in time with on/off ramps of 1.2 ns (see Section \ref{shape}) and pulse area equal to $2\pi$. The coupling to the environment is described in our model with the Lindblad master equation
\begin{equation}\label{Eq:master}
\frac{d\rho_\Delta}{dt}=-i[H,\rho_\Delta]+\sum_k [\hat{O}_k\rho_\Delta\hat{O}_k^+-\frac{1}{2}\{\hat{O}_k^+\hat{O}_k,\rho_\Delta\}],
\end{equation}
with the sum running over the following decoherence processes:
\begin{itemize}
\item spontaneous decay from $\ket{A_2}$ to $\ket{0}$ at the rate $\Gamma_0$ with the jump operator $O_0=\ket{0}\bra{A_2}$,
\item spontaneous decay from $\ket{A_2}$ to $\ket{-1}$ at the rate $\Gamma_{-1}$ with the jump operator $O_{-1}=\ket{-1}\bra{A_2}$,
\item spontaneous decay from $\ket{A_2}$ to $\ket{1}$ at the rate $\Gamma_1$ with the jump operator $O_1=\ket{1}\bra{A_2}$,
\item orbital dephasing of the level $\ket{A_2}$ at the rate $2\cdot\Gamma_{\phi}$ with the jump operator $O_{\phi}=\ket{A_2}\bra{A_2}$.
\end{itemize}
We numerically solve the Lindblad equation to obtain the final density matrix after the laser is switched off and all residual population has decayed from the excited $\ket{A_2}$ level. The decay rates were experimentally measured in a previous experiment on the same NV center at similar strain and magnetic field conditions \cite{Zhou2016Supp}. Their values are summarized in Table \ref{Decoherence rates}. The ground state decoherence times are significantly longer than the timescales of the experiment and thus can be neglected.

To further describe experimental losses, we include the effect of spectral hopping. This arises during NV center initialization by a 532 nm laser field, which is thought to excite the nearby charge environment. This creates some random electric field that slightly shifts the transition energy of $\ket{A_2}$ and thus changes the value of $\Delta$. We include this effect into the model by propagating the Lindblad equation for different values of $\Delta$, for which we assume a Gaussian distribution. Averaging over this distribution, the result for the final density matrix can be written as
\begin{equation}
\rho_{final}=\int_{-\infty}^{\infty}d\Delta\frac{\exp(-\frac{(\Delta-\Delta_0)^2}{2\sigma^2})}{\sqrt{2\pi}\sigma}\rho_{\Delta}^{final},
\end{equation}
where $\Delta_0$ is the intended detuning value. The standard deviation $\sigma$ is determined from best fit to the power dependence of the Pauli-$Z$ gate fidelity (Fig. 2e main text) as this gate requires only one laser frequency and thus is less prone to experimental imperfections. We find $\sigma$ to be $\sim$15  MHz.\\

 \begin{table}[h!]
\centering
\begin{tabular}{ c  c  c  c  c }
\hline
$\sigma/2\pi$ & $\Gamma_0/2\pi$ & $ \Gamma_{-1}/2\pi$ & $ \Gamma_{1}/2\pi$&  $\Gamma_{orb}/2\pi$\\
\hline
15 MHz & 1.6 MHz& 8.5 MHz & 4.3 MHz& 8.8 MHz\\
\hline
\end{tabular}
\caption{The values for the spectral hopping and the decoherence rates used to obtain the final density matrix after the excitation pulse.}
\label{Decoherence rates}
\end{table} 
\subsection{Calculation of fidelity of the gates and output states}
Every single-qubit gate can be characterized in terms of a quantum process 
\begin{equation}
\rho_{final}=\sum_{i,j=1}^{4}\chi_{ij} E_i\rho_{initial} E_j,
\end{equation}
where  $E_1=I$, $E_2=\sigma_x$, $E_3=-i\sigma_y$, $E_4=\sigma_z$ are given in terms of identity and Pauli matrices. The corresponding $\chi$-matrix completely describes the process and thus can be used to characterize the gates.

First, we consider the resonant gates with $\Delta=0$. Without decoherence and spectral hopping, the Hamiltonian described above would induce a rotation on the $\ket{\pm 1}$ Bloch sphere around the axis ($\theta,\phi$) by $\pi$ radians. This is the ideal gate that we seek to realize. Next, we analytically calculate the matrix $\chi_{ideal}$ for this ideal gate and extract the matrix $\chi_{sim}$ from our final density matrix obtained from numerical integration of Eq. \ref{Eq:master}. The fidelity of the gate is defined as $\rm{Tr}(\chi_{\emph{\text{sim}}}\chi_{\emph{\text{ideal}}})$. 

\begin{figure}[hb]
	\includegraphics[width=0.6\textwidth]{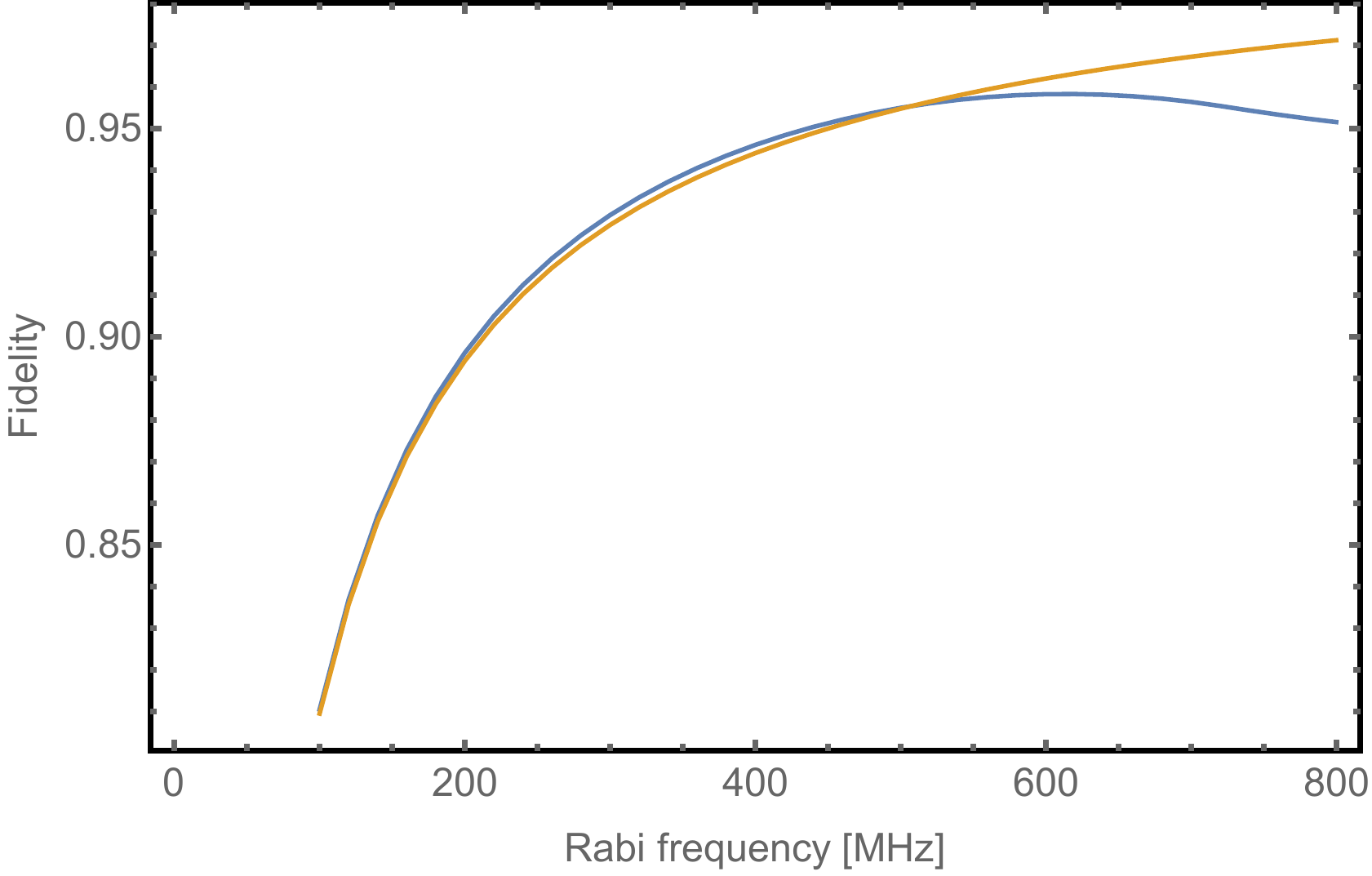}
	\caption{Dependence of the Y($\pi/2$) gate fidelity on the Rabi frequency $\Omega/2\pi$ at pulse maximum. The yellow curve corresponds to the rectangular pulse; the blue one to the experimental pulse shape (trapezoid with 1.2 ns ramp on/off). Above $\Omega/2\pi \approx 600$ MHz, pulse shape errors begin to limit the attainable fidelity.}
	\label{Pulse_form}
\end{figure}

For a non-resonant gate, the situation is more complicated, as even without the decoherence and spectral hopping, the Hamiltonian described above may not generate an ideal gate on the logical subspace for non-perfect square pulses \cite{Sjoqvist2016Supp}. It is possible to show that some residual population may still remain in the $\ket{A_2}$ excited state after the pulse is turned off. This effect decreases the gate fidelity because the ideal gate is defined in the absence of decoherence, spectral hopping, and errors to pulse shape or area. In the ideal case, our gate would be a counter-clockwise rotation around the axis ($\theta, \phi$) by the angle $\gamma=\pi(1-\Delta/\sqrt{\Delta^2+\Omega^2})$. We can analytically calculate $\chi_{ideal}$ for this gate and then compare it to $\chi_{sim}$ obtained from the numerical simulation, which includes the effects of both decoherence and the non-ideal pulse shape. These effects can be considered separately to determine relative contributions. Figure \ref{Pulse_form} compares the fidelities of the $Y(\pi/2)$ gates if we use an ideal rectangular pulse (blue) or the experimental trapezoid pulse (orange). Here we plot the fidelity versus Rabi frequency at the maximum of the pulse, $\Omega/2\pi$. One can infer that below 600 MHz the fidelity is mostly limited by spectral hopping and decoherence and the effect of the pulse form is negligible.

Finally, we consider the initial state $\ket{\psi_i}$ that we would like to transform into the final state $\ket{\psi_f}$ using the gate $U$. Due to decoherence and spectral hopping, we will end up at some mixed state with the density matrix $\rho_f$. We then define the fidelity with which we can transform the state $\ket{\psi_i}$ into the state $\ket{\psi_f}$ as Tr$(\ket{\psi_f}\bra{\psi_f} \rho_f)$. These values are used to compare with the experimental projections and their derived Bloch vector amplitudes (e.g., Fig. 3a,b and 4a of main text).

\end{document}